\documentclass[12pt]{article}
\usepackage{amsmath}
\usepackage{amssymb}
\begin{document}
\vskip 2cm
\begin{center}
{\sf{\Large Interior of Schwarzschild black hole as a relativistic free particle}}

\vskip 2cm

{\sf {\it\bf Dheeraj Shukla}}\\
{\it Physics Department, Center of Advanced Studies,}\\
{\it Banaras Hindu University, Varanasi - 221 005, (U.P.), India}\\

\vskip 0.1cm

\vskip 0.1cm

\small {\bf E-mail: {\it dheerajkumarshukla@gmail.com}}

\end{center}

\vskip 2cm

\noindent
{\bf Abstract:} Using the standard Hamiltonian approach followed by ADM formalism to the geometry of a Schwarzschild black hole, we show that the interior spacetime of Schwarzschild black hole behaves like a relativistic free particle with Planck mass $M_{p}$. We further present the (anti-)BRST quantization of the relativistic free particle representing the interior of Schwarzschild black hole. With help of the (anti-)BRST formalism, we determine the physical states of the system in general Hilbert space. This is a novel observation in the context of gauge theoretic approach to gravity.\\



\newpage
\section{Introduction}

The most familiar interaction, gravity is described by Einstein's general relativity theory. Gravity is a classical theory and deals with the underlying spacetime geometry. On the other hand, the gauge theory which describes the rest of three fundamental interactions of nature, is a quantum theory. This has been a long awaited outstanding problem to provide a quantum description of gravity or a spacetime geometric version of gauge theories. There have been made many attempts to provide a quantum picture of the gravity e.g. string theory, loop quantum gravity etc. To quantize the gravity one needs to rewrite the Einstein's equations in Hamiltonian or Lagrangian formulation, best suited for quantum theories. The Hamiltonian formulation of gravity, was proposed firstly by P. A. M. Dirac \cite{PAMD}. This approach got major attention in community and opened a new regime in theoretical physics. Gravity is a large scale interaction while the gauge interactions are small scale interactions; therefore, a theory which incorporates the gravity and quantum mechanics both, should meet the extreme conditions of spacetime scaling. A black hole is such an object where the underlying spacetime is so warped that a quantum description of gravity becomes inevitable.

Black hole physics has been one of the leading area in theoretical physics since the derivation of first non-trivial solution of Einstein's equation for a homogeneous and isotropic spacetime by Karl Schwarzschild in his famous paper \cite{EM}. This attracted the attention of many leading physicists in course of time. The most fascinating point of the Schwarzschild black hole is the concept of event horizon; a hypersurface around the Black hole which acts like a boundary to it. Even the fastest moving light rays, can not escape from there if once entered in the interior of the boundary. There are many such bizarre things associated with the event horizon of the black holes. One such a well known result for the interior of a Schwarzschild black hole, is the interchange of behavior of space and time coordinates. The time behaves as if it were the radial coordinate while the radial coordinate behaves as if it were a time coordinate, a monotonically increasing sequence of real numbers \cite{Inv., MTW}.

\section{Developments}

The general spherically symmetric and isotropic spacetime metric \cite{SB} in Minkowskian 4-dimensions can be written as,
\begin{equation}
{ds}^2 = -B(r) \,dt^2 + \frac{N^2(r)}{B(r)}\, dr^2 + r^2 d\Omega^2
\end{equation}
\label{eq1}
where $ B(r) $ is a non-zero quantity and
\begin{eqnarray}
d\Omega^2 = d\theta^2 +  sin^2\theta \,  d\phi^2
\end{eqnarray}
\label{eq2}
and each metric component is dependent on the radial coordinate $r$. A substitution for $N$ and $B$
\begin{equation}
B(r) = 1- \frac{2 M G}{r}, \qquad N(r) = 1
\end{equation}
\label{eq3}
can convert the metric into the well known Schwarzschild metric \cite{EM, Inv.} as,
\begin{equation}
ds^2 = - \left( 1- \frac{2M G}{r}\right)\, dt^2 + \left(1- \frac{2 M G}{r} \right)^{-1}\, dr^2 + r^2\, d\Omega^2
\label{eq4}
\end{equation}
where  $M$ is the mass of the spherically symmetric object (Schwarzschild BH in our case).
Here, $r= 0$ is the real singularity while $r= 2GM  $ is a coordinate singularity. By suitable coordinate transformations, we can eliminate the coordinate singularity and then $r = 2GM $ becomes the event-horizon of the spherically symmetric object (black hole). It has been established that inside the event horizon, the metric component  $g_{tt}$ turns out to be positive and $g_{rr}$ turns out to be negative. Thus, the role of temporal and radial coordinates are mutually interchanged. We can express this situation as
\begin{equation}
t\rightarrow r, \quad r\rightarrow t, \quad N(r)\rightarrow N(t), \quad B(r)\rightarrow - B(t)
\end{equation}
\label{eq5}
with this, the metric inside the event-horizon \cite{SB, KKTY}, turns out to be
\begin{equation}
ds^2 = - \frac{N^2(t)}{B(t)}\, dt^2 + B(t)\, dr^2 + H^2(t)\,d\Omega^2
\end{equation}
\label{eq6}
where $H(t)$ takes care of the function of $ r $  converting into a function of $ t $ and thus bears the information of the polar geometry of the Black-hole.
The vacuum Hilbert-Einstein action is
\begin{equation}
{\cal S} =  \frac{1}{16 \pi G}\,\int d^{4}x\,\sqrt{-g^{(4)}}\,  R^{(4)}
\end{equation}
\label{eq7}
Here $ g^{(4)} $ is the determinant of 4-dimensional metric given by Eq.(6) and $ R^{(4)} $ is the 4-dimensional Ricci curvature scalar.
Now, writing the same action in well known $(3+1)$ ADM formalism \cite{MTW, KKTY} as,
\begin{equation}
{\cal S} = \int dt\,L
\end{equation}
\label{eq8}
where the Lagrangian is given as\cite{Inv., SB}
\begin{eqnarray}
L = \frac{1}{16}\, \int d^{3}x \,N \sqrt{g^{(3)}}\, \left( K_{ab}K^{ab} - K^{2} + R^{(3)}\right)
\end{eqnarray}
\label{eq9}
Here $ N $ is the lapse function and is actually related to the metric component $ g_{tt} $  and $ K_{ab} $ denotes the extrinsic curvature, $ K $ is the trace of $ K_{ab} $ while $ g^{(3)} $ and $ R^{(3)} $ denote the determinant and the Ricci curvature scalar for the 3-dimensional spatial metric respectively. Inserting the metric (6) in the expression for the Lagrangian (9), the Lagrangian reduces to the form \cite{SB, KKTY}
\begin{equation}
L =  \frac{V_0}{8 \pi G} \left[ N - \frac{1}{N}(H \dot H \dot B + \dot H^2 B) \right]
\end{equation}
\label{eq10}
where $V_0 = 4\pi \int dr $ and $ \dot{B} = \frac{d}{dt}B $.
Above Lagrangian can be further rewritten as,
\begin{equation}
L =  V_{0}\, M^{2}_{p}\/\left[N - \frac{1}{N}(H \dot H \dot B + \dot H^2 B)\right].
\end{equation}\label{eq11}
where $ M_{p} $ is the Planck mass and is given as,
\begin{equation}
M^{2}_{p} = \frac{c\,\hbar}{8 \pi G}.
\end{equation}
\label{eq12}
In natural units $ c=1,\,  \hbar =1 $, the expression for the Planck mass becomes,
\begin{equation}
M^{2}_{p} = \frac{1}{8 \pi G}.
\end{equation}
\label{eq13}
Now, let us substitute the following,
\begin{eqnarray}
x - iy &=& H,\quad x + iy = HB, \nonumber\\
\frac{dH}{dt} &=& \dot{H} = \dot{x} - i\dot{y}
\end{eqnarray}
\label{eq14}
and also,
\begin{eqnarray}
\frac{d}{dt}(HB) = \frac{d}{dt} (x+\,iy) = \dot{x} + i\dot{y}, \nonumber\\
\dot{H} B + H \dot{B} = \dot{x} + i\dot{y}.
\end{eqnarray}
\label{eq15}
where $ i = \sqrt{-1} $.
Multiplication of equation (14) with equation (15) gives,
\begin{eqnarray}
\dot{H}^2 B + H \dot{H} \dot{B} = \dot{x}^2 + \dot{y}^2.
\end{eqnarray}
\label{eq16}
Substituting the above expression (16) in the expression of Lagrangian (11), we obtain,
\begin{equation}
L =  M^{2}_{p}\,V_{0} \left[ N - \frac{1}{N}(\dot{x}^2 + \dot{y}^2) \right].
\end{equation}
\label{eq17}

\section{Interior: A model of relativistic free particle}

Now, in order to make the expression (17) more simplified, let us substitute the expressions fro x and y in a parametric form,
\begin{equation}
x = \frac{i}{M_{p}}\, cos\, \varphi, \quad y = \frac{i}{M_{p}}\, sin\,\varphi.
\end{equation}
\label{eq18}
where $ \varphi $ is a scalar quantity and can be calculated in terms of dynamical quantity 
$ B $ as,
\begin{equation}
\varphi = tan^{-1}\left (i\,\frac{1-B}{1+B}\right).
\end{equation}
\label{eq19}
With these substitutions the Lagrangian (17) becomes,
\begin{equation}
L =  V_{0}\,\left[  \frac{\dot{\varphi}^2}{N} + N\,M^{2}_{p}\right] .
\end{equation}
\label{eq20}
This Lagrangian has exactly the same form as the 1-dimensional Lagrangian for relativistic free particle \cite{BVH, RPM} with mass $ M_{p} $ apart from a factor $ V_{0} $. This factor doesn't alter the dynamics of the system at all. Surprisingly, the dynamics of the underlying spacetime components is such that spacetime itself behaves like a relativistic free particle of Planck mass $ M_{p} $, moving in flat spacetime.
The above Lagrangian is endowed with first-class constraint as it is obvious from the absence of the canonical momentum for $ N $. The primary constraints in the theory is
\begin{equation}
\Pi_{N} \approx 0 = \Omega_{1}.
\end{equation}
\label{eq21}
The corresponding momentum for $ \varphi $  can be calculated as follows,
\begin{equation}
\Pi_{\varphi} = \frac{2\,V_{0}}{N}\, \dot{\varphi}.
\end{equation}
\label{eq22}
The canonical Hamiltonian therefore can be calculated as follows,
\begin{equation}
H_{can} = \dot{\varphi}\,\Pi_{\varphi} - L.
\end{equation}
\label{eq23}
which turns out to be in the form given below when the values of  $ \Pi_{\varphi} $ and $ L $ are substituted in above expression,
\begin{equation}
H_{can} = N \left( \frac{\Pi^{2}_{\varphi}}{4V_{0}}\, - V_{0}\, M^{2}_{p}\right).
\end{equation}
\label{eq24}
The primary Hamiltonian can be calculated this way,
\begin{equation}
H_{p} = H_{can} + \xi\, \Pi_{N}.
\end{equation}
\label{eq25}
where $ \xi $ is the Lagrange's multiplier and turns out to be a scalar quantity.
With proper substitutions the primary Hamiltonian can be found to be,
\begin{equation}
H_{p} = N \left( \frac{\Pi^{2}_{\varphi}}{4V_{0}}\, - V_{0}\, M^{2}_{p}\right) + \xi \, \Pi_{N}.
\end{equation}
\label{eq26}
Now the secondary constraint of the theory can be calculated using the formula,
\begin{equation}
\Omega_{2} = \frac{d}{dt}\Omega_{1} = \frac{d}{dt} \Pi_{N} = \left\lbrace \Pi_{N}, H_{p} \right\rbrace _{PB}
\end{equation}
\label{eq27}
where $\left\lbrace..,.. \right\rbrace_{PB} $ is nothing but the classical Poisson bracket.
Various canonical non-vanishing possible Poisson brackets are listed below in this case,
\begin{equation}
\left\lbrace \varphi, \Pi_{\varphi} \right\rbrace_{PB} =1,\quad \left\lbrace N, \Pi_{N} \right\rbrace_{PB} = 1.
\end{equation}
\label{eq28}
With above set of Poisson brackets, one can calculate the secondary constraint on the Lagrangian as follows,
\begin{equation}
\Omega_{2} = V_{0}\, M^{2}_{p} - \frac{\Pi_{\varphi}^{2}}{4V_{0}} \approx 0.
\end{equation}
\label{eq29}
Using the same method one can make sure that there is no further constraint on the theory. And the Dirac classification of constraints \cite{Dir, HT} tells us that both the primary and secondary constraints on the Lagrangian are first-class in nature. Therefore, the gauge generator for the corresponding gauge transformations can be calculated using the formula \cite{Dir,HT},
\begin{equation}
{\cal G} = \dot{\xi}\, \Omega_{1} - \xi\, \Omega_{2},
\end{equation}
\label{eq30}
or
\begin{equation}
{\cal G} =  \dot{\xi}\,\Pi_{N} + \xi\, \left( \frac{\Pi^{2}_{\varphi}}{4V_{0}} -  V_{0}\,M_{p}^{2} \right).
\end{equation}
\label{eq31}
The gauge transformations can be calculated in the following manner,
\begin{eqnarray}
\delta_{g} \varphi &=& \left\lbrace \varphi, {\cal G } \right\rbrace_{PB} =\, \frac{\xi\, \dot{\varphi}}{N}, \nonumber\\
\delta_{g} N &=& \left\lbrace N, {\cal G} \right\rbrace_{PB} = \,\dot{\xi}.
\end{eqnarray}
\label{eq32}
Under this symmetry transformations the Lagrangian (20) varies as,
\begin{equation}
\delta_{g}\,L = \frac{d}{dt} \left[ \left( \frac{\dot{\varphi}^2}{N^2} + M^{2}_{p}\right)\xi\, V_{0} \right].
\end{equation}
\label{eq33}
A total time derivative. This ensures us again that the action remains quasi-invariant under the symmetry transformations given by equation (32). According to Noether's theorem, there must be a  conserved current for this set of symmetry transformations. The Noether conserved current thus turns out to be
\begin{eqnarray}
Q_{g} = \xi \,V_{0}\left(\frac{\Pi_{\varphi}^2}{4V_{0}^2} - M^{2}_{p} \right)  + \dot{\xi}\, \Pi_{N}.
\end{eqnarray}
\label{eq34}
For one dimensional case the conserved current and conserved charge $Q_{g}$ both are same.
Now, under the application of equations of motion,
\begin{equation}
\ddot{\varphi} = \frac{\dot{\varphi}\,\dot{N}}{N},
\end{equation}
\label{eq35}
\begin{equation}
{\dot{\varphi}}^2 = M^{2}_{p}\, {N}^2.
\end{equation}
\label{eq36}
The charge $Q_{g}$ can be easily shown to be conserved. This conserved charge itself is able to generate all the symmetry transformations, given by equation(32), with use of the formula,
\begin{equation}
\delta_{g} \Phi =   \lbrace\Phi, Q_{g}\rbrace.
\end{equation}
\label{eq37}
where $\Phi= \varphi, \tilde{N}$.

\noindent

\section{Nilpotent (Anti-)BRST Symmetries in Lagrangian Formulation}

The Lagrangian (20) is gauge invariant but is endowed with first-class primary constraint,
\begin{equation}
\Pi_{N} \approx 0.
\end{equation}
\label{eq38}
It can be shown that with a gauge-fixing term ($- \frac{1}{2}\, \dot{N}^2 $), one can form a modified Lagrangian with no constraints, i.e.
\begin{equation}
L =    V_{0}\,\left(  \frac{\dot{\varphi}^2}{N}  + N\,M^{2}_{p}\right) - \frac{1}{2}\,\dot{N}^{2}.
\end{equation}
\label{eq39}
This Lagrangian is free from the constraints and hence suitable for the canonical quantization. But this comes at the cost of gauge invariance of the theory. The gauge invariance is gone due to gauge-fixing term. To restore the gauge invariance in the Lagrangian (39), we add an extra Faddeev-Popov (anti-)ghost term $ i \dot{\bar{C}}\dot{C}$ to the modified Lagrangian (39). The new gauge-fixed (anti-)BRST Lagrangian \cite{SKRPM} thus becomes,
\begin{equation}
L_{b}  =  V_{0}\,\left(  \frac{\dot{\varphi}^2}{N}  + N\,M^{2}_{p}\right) - \frac{1}{2}\,\dot{N}^{2} + i \,\dot{\bar{C}}\dot{C}.
\end{equation}
\label{eq40}
where the (anti-)ghost terms $ C $ and $ \bar{C} $ are subjected to following conditions:
\begin{equation}
C^{2} =0 = {\bar{C}}^{2}, \quad C\bar{C} + \bar{C}C = 0.
\end{equation} The  BRST symmetry transformations for above Lagrangian, are as follows,
\begin{equation}
\label{eq41}
s_b\, \varphi = \frac{\dot{\varphi}}{N}\,C,\quad s_b\,N =  \dot{C},\quad s_b\, {C} = 0,\quad s_b\, \bar{C} = i\, \dot{N}.
\end{equation}
\label{eq42}
The Lagrangian (40), under the set of BRST transformations (41) varies as,
\begin{equation}
s_b\, L_{b} = \frac{d}{dt} \left[ \left(  \frac{{\dot{\varphi}}^2}{N^2} + M^{2}_{p} \right)C\,V_{0} -\dot{N}\dot{C}\right].
\end{equation}
\label{eq43}
A total time derivative which ensures the action to be quasi-invariant under above BRST symmetry transformations. In a similar fashion it can be shown that the Lagrangian (40) varies under the set of anti-BRST transformations,
\begin{equation}
s_{ab}\,\varphi = \frac{\dot{\varphi}}{N}\,\bar{C},\quad s_{ab}\,N =  \dot{\bar{C}},\quad s_{ab}\, \bar{C} = 0,\quad s_{ab}\, C = -i\, \dot{N}.
\end{equation}
\label{eq44}
as
\begin{equation}
s_{ab}\,  L_{b} = \frac{d}{dt} \left[ \left(  \frac{{\dot{\varphi}}^2}{N^2} + M^{2}_{p} \right)\bar{C}\,V_{0} -\dot{N}\dot{\bar{C}}\right].
\end{equation}
\label{eq45}
It is easy to check that the on-shell nilpotency of order-2 and the on-shell anti-commutativity are the essential features of these sets of (anti-)BRST transformations i.e.
\begin{equation}
s_b^2 = 0 = s_{ab}^2, \qquad s_b\, s_{ab} + s_{ab}\,s_b = 0.
\end{equation}
\label{eq46}
Corresponding to these symmetry transformations, there exist conserved Noether charges which are nothing but the conserved currents themselves. In order to calculate the conserved charges of the BRST and anti-BRST symmetries as well as, we need to calculate the canonical momenta first, which are
\begin{equation}
\Pi_\varphi = \frac{2\,V_{0}}{N}\,\dot{\varphi},\quad \Pi_{N} = - \dot{N},\quad \Pi_C = -i\, \dot{\bar{C}}, \quad \Pi_{\bar{C}} = i\, \dot{C}.
\end{equation}
\label{eq47}
The various equations of motion for the Lagrangian (40) are following:
\begin{eqnarray}
\ddot{\varphi} &=& \frac{\dot{\varphi} \dot{N}}{N},  \nonumber\\
\ddot{N} &=& \frac{V_{0}\,\dot{\varphi}^{2}}{N^{2}} - V_{0}\, M^{2}_{p},\nonumber\\
\ddot{\bar{C}} &=& 0, \nonumber\\
\ddot{C} &=& 0.
\end{eqnarray}
\label{eq48}
The Noether conserved current and hence the conserved charges corresponding to BRST and anti-BRST symmetry transformations can be calculated easily and are found to be
\begin{eqnarray}
Q_b &=& \frac{\Pi^{2}_{\varphi}}{4 V_{0}}\,C - M^{2}_{0}\,V_{0}\,C- i\, \Pi_{N}\,\Pi_{\bar{C}}, \nonumber\\
Q_{ab} &=& \frac{\Pi^{2}_{\varphi}}{4 V_{0}}\,\bar{C} - M^{2}_{0}\,V_{0}\,\bar{C} + i \,\Pi_{N}\,\Pi_{C}.
\end{eqnarray}
\label{eq49}
Further, using the various equations of motion (47), this can be shown that both the charges are invariant in time. Now, the operator corresponding to various dynamical variables in Lagrangian (40) under canonical quantization scheme produce the following (anti-)commutators,
\begin{equation}
[\hat{\varphi},\, \hat{\Pi_{\varphi}}] = i\hat{I}, \quad [\hat{N}, \,\hat{\Pi}_{N}] = i\hat{I},\quad \lbrace C,\, \Pi_{C}\rbrace = i\hat{I},\quad \lbrace {\bar{C}},\, \Pi_{\bar{C}}\rbrace = i\hat{I}.
\end{equation}
\label{eq50}
where $ \hat{I} $ is the Identity operator.

We have taken the natural units $c = 1 $ and $\hbar = 1$ everywhere.
The (anti-)commutators (50) can be obtained in terms of our usual field variables as,
\begin{equation}
[\hat{\varphi}, \,{\hat{\dot{\varphi}}}]= i\,\frac{\hat{N}}{2V_{0}},\quad [\hat{N},\,{\hat{\dot{N}}}] = -i\hat{I} ,\quad \lbrace C,\,\dot{\bar{C}}\rbrace = -\hat{I},\quad \lbrace\bar{C},\, \dot{C}\rbrace = \hat{I}.
\end{equation}
\label{eq51}
On the basis of (anti-)commutation relations (50) the absolute anti-commutativity and nilpotency of order-2 of the (anti-)BRST charges can be established, i.e.
\begin{eqnarray}
Q^2_b &=& 0  = \quad Q^2_{ab}, \nonumber\\
\lbrace Q_b,\, Q_{ab}\rbrace &=& Q_b Q_{ab} + Q_{ab}Q_b  =\quad 0.
\end{eqnarray}
\label{eq52}
With all these treatise one can easily check that the charges $Q_b$ and $Q_{ab}$ are able to generate all the BRST and anti-BRST transformations respectively by using the formula
\begin{eqnarray}
s_b \,\chi &=&  \pm\, i\,\left[  \chi,  \, Q_b \right]_{\pm}, \nonumber\\
s_{ab}\,\chi &=& \pm\, i\,\left[ \chi,  \, Q_{ab} \right]_{\pm}.
\end{eqnarray}
\label{eq53}
where $\chi = \hat{\varphi}, \hat{N}, C, \bar{C}$
and  for some generic fields $ A $ and $ B $, we have,
$ [A, B]_{+}= \lbrace A,B \rbrace $ and $ [A, B]_{-} = [A, B] $.
Further, the physicality criteria defines the physical states of the system in the corresponding quantum Hilbert space. The physicality criteria picks only those states to be physical which are annihilated by the conserved Noether charges, i.e. $ Q_{a(b)}\mid phys> \,=\, 0 $. Thus, the physical states are those, for which,
\begin{eqnarray}
Q_{b}\mid phys> \,=\, 0 \Rightarrow \left( \frac{\Pi^{2}_{\varphi}}{4 V_{0}}\,C - M^{2}_{0}\,V_{0}\,C- i\, \Pi_{N}\,\Pi_{\bar{C}}\right) \mid phys> \,=\, 0, \nonumber\\
Q_{ab}\mid phys> \,= \,0 \Rightarrow \left( \frac{\Pi^{2}_{\varphi}}{4 V_{0}}\,\bar{C} - M^{2}_{0}\,V_{0}\,\bar{C} + i \,\Pi_{N}\,\Pi_{C}\right) \mid phys> \,= \,0.
\end{eqnarray}
\label{eq54}
consistent with Dirac's prescription of constrained systems \cite{Dir}, \cite{HT}. Now, substituting the corresponding operator forms for $\hat{\varphi}$ and $\hat{N}$ in the (anti-)commutators given by equation (51), we obtain the commutation relations among the metric components of the spherically symmetric and isotropic metric (6), as follows,
\begin{eqnarray}
\left[ ln \hat{B(t)},\, \frac{d}{dt}(ln \hat{B(t)}) \right]   &=& -2 i\, \frac {\hat{N(t)}}{V_{0}},\nonumber\\
\left[ \hat{N(t)}, {\hat{\dot{N(t)}}}\right]   &=& -i\hat{I}.
\end{eqnarray}
\label{eq55}
But this result is valid for the interior of the Schwarzschild black hole. The worth observation for us is the observation taken far from the interior of the Schwarzschild black hole. Fortunately, we can reverse the transformation which actually got interchanged in the interior. The appropriate transformations, far away from the event horizon are:
\begin{equation}
t\rightarrow r,\quad r\rightarrow t,\quad,B(t)\rightarrow B(r),\quad N(t)\rightarrow -N(r).
\end{equation}
\label{eq56}
With this set of transformations, the commutation relations (55) turns out to be,
\begin{eqnarray}
\left[ ln \hat{B}(r),\, \frac{d}{dr}ln \hat{B}(r) \right]   &=& 2 i\, \frac {\hat{N}(r)}{V_{0}},\nonumber\\
\left[ \hat{N}(r), \frac{d}{dr}\hat{N}(r)\right]   &=& i\hat{I}.
\end{eqnarray}
\label{eq57}
These commutation relations  infer that the lapse function $ N(r) $ and its gradient $ \frac{d}{dr}N(r) $, both can not be known simultaneously. And similarly, the values of logarithm of the radial component of the metric $ ln B(r) $ and its gradient $ \frac{d}{dr}ln B(r) $ can not be measured simultaneously.

\noindent
\section{Conclusions and Discussions}

The central theme of our endeavor has been accomplished. The most fascinating consequence of our work is that a highly warped spacetime interior of the Schwarzschild black hole, behaves like a relativistic free particle in flat Minkowskian spacetime. This is a {\it duality} between gauge sector and gravitation sector. Thus, we have been able to replace the whole curved geometry without any dynamical force involved, by a well known and easily solvable problem of relativistic free particle in flat Minkowskian spacetime.  The mass of this particle which is none the else but the chunk of local spacetime, turns out to be Planck mass. We have also been able to provide a complete (anti-)BRST and usual gauge theoretic quantization of the interior of the Schwarzschild black hole. Furthermore, we have also been able to define the physical states out of the total Hilbert space, for the interior of the Schwarzschild black hole. Along with this, we have extracted out the information that the lapse function  $ N(r) $ and its gradient both can not be determined simultaneously. And the same is also true for the logarithm of the radial metric component $ B(r) $ and its gradient. Furthermore, we are able to show that the lapse function is connected to the radial component of the metric. 

Our next plan is to extend this work to explore the connection between a Kerr black hole and the relativistic spinning free particle. And thus, to show that the whole black hole dynamics can be mapped into a corresponding particle dynamics in flat Minkowskian spacetime.\\

\vskip 1cm

\noindent
{\bf Acknowledgements:}\\

I am very grateful to Prof. Loriano Bonora from Trieste, Italy (who visited our group at the physics department for a couple of weeks in December, 2013) for his valuable suggestions and discussions in the problem. I am also very thankful to my supervisor Prof. R. P. Malik for his generous support and encouragement for working on this problem. I would like to thank UGC, Government of India, New Delhi, for financial support through RFSMS scheme (F. 5-127/2007 (BSR), March 2013).\\

\end{document}